\pdfoutput=1

\documentclass[hidelinks,12pt]{article}

\usepackage{hyperref, url} 
\usepackage{graphicx,amsfonts,psfrag,layout,subcaption,array,longtable,lscape,booktabs,dcolumn,amsmath,amssymb,amssymb,amsthm,setspace,epigraph,chronology,color,colortbl,wasysym,diagbox,natbib,colortbl,authblk,commath,upgreek,bbm,slashbox,threeparttable,bm,comment,adjustbox}
\usepackage[]{graphicx}\usepackage[]{color}
\usepackage[page]{appendix}
\usepackage[section]{placeins}
\usepackage[linewidth=1pt]{mdframed}
\usepackage[margin={1in}]{geometry} 

\vspace{-2cm}

\title{Retrospective causal inference via matrix completion, with an evaluation of the effect of European integration on cross-border employment\thanks{\emph{Acknowledgments}: Poulos and Li's research is partially supported by the National Science Foundation under Grant DMS-1638521 to the Statistical and Applied Mathematical Sciences Institute. Albanese's research is supported by the Luxembourg National Research Fund (C17/SC/11700060). The views expressed herein are those of the authors and do not necessarily reflect those of the institutions they represent. Code and microdata access instructions are available at: \url{https://github.com/jvpoulos/schengen}.}}
\author[1]{Jason Poulos\thanks{\emph{Address for correspondence:} 180 Longwood Avenue, Boston, MA 02115. \emph{E-mail:} poulos@hcp.med.harvard.edu.}}
\author[2]{Andrea Albanese}
\author[3]{Andrea Mercatanti}
\author[4]{Fan Li}
\affil[1]{Department of Health Care Policy, Harvard Medical School}
\affil[2]{Luxembourg Institute of Socio-Economic Research (LISER)}
\affil[3]{Bank of Italy}
\affil[4]{Duke University}

\date{}
\usepackage{xr}
\usepackage[bottom]{footmisc}

\usepackage{mathtools,tikz,caption}
\DeclareRobustCommand\sampleline[1]{%
	\tikz\draw[#1] (0,0) (0,\the\dimexpr\fontdimen22\textfont2\relax)
	-- (2em,\the\dimexpr\fontdimen22\textfont2\relax);%
}

\usetikzlibrary{plotmarks}

\usepackage{multicol}

\usepackage{xcolor}

\definecolor{Darjeeling11}{HTML}{FF0000}
\definecolor{Darjeeling14}{HTML}{F98400}
\definecolor{Darjeeling15}{HTML}{5BBCD6}

\definecolor{lightblue}{HTML}{5eb1ff}
\definecolor{darkblue}{HTML}{006eb3}
\definecolor{lightgreen}{HTML}{65c97b}

\usepackage{footmisc}
\DefineFNsymbols{mySymbols}{{\ensuremath\dagger}{\ensuremath\ddagger}\S\P
   *{**}{\ensuremath{\dagger\dagger}}{\ensuremath{\ddagger\ddagger}}}
\setfnsymbol{mySymbols}

\renewcommand\footnotemark{}

\usepackage{tabularx}
\newcolumntype{Y}{>{\raggedleft\arraybackslash}X}

\definecolor{Gray}{gray}{0.9}

\newcommand{\captionfonts}{\normalsize}

\makeatletter  
\long\def\@makecaption#1#2{%
  \vskip\abovecaptionskip
  \sbox\@tempboxa{{\captionfonts #1: #2}}%
  \ifdim \wd\@tempboxa >\hsize
    {\captionfonts #1: #2\par}
  \else
    \hbox to\hsize{\hfil\box\@tempboxa\hfil}%
  \fi
  \vskip\belowcaptionskip}
 
\newtheorem*{assumption*}{\assumptionnumber}
\providecommand{\assumptionnumber}{}


\newcommand{\diag}{\mathrm{diag}}

\DeclareMathOperator*{\argmin}{arg\,min}

\newcommand{\bY}{\mathbf{Y}}

\DeclareMathOperator\bE{\mathbb E} 

\begin{document} 

\begin{singlespacing}
\maketitle  
\end{singlespacing}
\thispagestyle{empty}
%
 \vspace{-10mm}
\begin{abstract}  
\noindent
We propose a method of retrospective counterfactual imputation in panel data settings with later-treated and always-treated units, but no never-treated units. We use the observed outcomes to impute the counterfactual outcomes of the later-treated using a matrix completion estimator. We propose a novel propensity-score and elapsed-time weighting of the estimator's objective function to correct for differences in the observed covariate and unobserved fixed effects distributions, and elapsed time since treatment between groups. Our methodology is motivated by studying the effect of two milestones of European integration---the Free Movement of persons and the Schengen Agreement---on the share of cross-border workers in sending border regions. We apply the proposed method to the European Labour Force Survey (ELFS) data and provide evidence that opening the border almost doubled the probability of working beyond the border in Eastern European regions.
\begin{singlespace}
\end{singlespace} 
\end{abstract}	

\pagebreak
\pagenumbering{arabic}

\section{Introduction} \label{intro}

We consider the problem of estimating the causal effects of a policy intervention in panel data settings where there exists units exposed to the policy after a given time (\emph{later-treated}) and units that are exposed to the policy throughout the study period (\emph{always-treated}). This is distinct from the standard setting in comparative case studies which consist of later-treated and never-treated units, but no always-treated units. Motivated by the setting with later-treated and always-treated units, we formulate a new identification strategy of \emph{retrospective} counterfactual imputation to predict potential outcomes under treatment for the later-treated in the pre-treatment period. We treat these potential outcomes as missing and impute them via the matrix completion method \citep{athey2017matrix}. We extend the matrix completion estimator two ways: first, by propensity- and elapsed-time weighting the objective function to correct for differences in the observed covariate and unobserved fixed effects distributions and elapsed time since treatment between treated and control groups; second, we explicitly include time-varying covariates in the outcome model and propose a procedure for imputing endogenous covariate values when reconstructing the missing potential outcomes.

A standard method for causal inference with panel data is difference-in-differences. This method hinges on the \emph{parallel trends} assumption, which states that in the absence of treatment, the average outcomes of treated and control units would have followed parallel paths. Under parallel trends, difference-in-differences identifies causal effects by contrasting the change in outcomes pre- and post-treatment, among the treated and control groups \citep{Ashenfelter1978,Ashenfelter1985,Card1994,Heckman1997,Callaway2018}. Estimation in difference-in-differences is traditionally tied with an additive fixed-effects regression model \citep{angrist2009}. Less model-dependent semiparametric approaches \citep{abadie2005semiparametric,ding2019bracketing}, as well as matching and weighting methods \citep{hazlett2018trajectory,kim2018matching,strezhnev2018semiparametric} have also been developed. However, the parallel trends assumption is generally invalid in the presence of unobserved time-varying confounders. In the retrospective setting, difference-in-differences identifies the effect of the treatment for later-treated in the pre-treatment period, but under a parallel trends assumption for the potential outcome under treatment. The matrix completion estimator relaxes this parallel trends assumption by allowing the presence of unobserved time-varying confounders.

A popular alternative method of handling unobserved time-varying confounders in panel data is the synthetic control method  \citep{abadie2003economic, abadie2010synthetic, abadie2015comparative}, which constructs a ``synthetic control unit'' as the counterfactual for single treated unit by reweighting the control units given their pre-treatment covariates or lagged outcomes. The idea is that conditioning on pre-treatment observables may help balance unobserved time-varying confounding between treatment and control groups. Recent methods have been developed to combine features of difference-in-differences and synthetic control \citep{arkhangelsky2019synthetic,ben2018augmented}, or perform dimension reduction prior to reweighting the control units using lasso regression or matrix factorization \citep{belloni2017program,carvalho2018arco,amjad2018robust}. \citet{doudchenko2016balancing} propose a generalization of that relaxes the convexity restriction of the synthetic control method, which precludes negative and nonlinear interactions between control unit outcomes. 

\citet{xu2017generalized} further generalize by allowing for multiple treated units and treatment periods using an interactive fixed effects (i.e. latent factor) model \citep{bai2003inferential,bai2009panel,gobillon2016regional}. The advantages of matrix completion over the interactive fixed effects models are three-fold: first, matrix completion does not require fixing the rank (i.e., number of unobserved factors) of the underlying data; second, it works in settings with staggered adoption of treatment \citep{athey2018design}; third, matrix completion uses all observed data to estimate unobserved factors, while interactive fixed effects models use pre-treatment data in the standard setting. An implication of the latter advantage is that matrix completion works well under weaker conditions; i.e, when there are a number of weak patterns over time that are common to units.  
%

Our proposed method is motivated by studying the effects of two milestones of the European integration, namely the Schengen Agreement and the Free Movement of persons, on the labour market of bordering regions. The Schengen Agreement introduced a common visa policy and abolished border controls between countries with the goal of boosting economic integration by facilitating crossing of borders. Removing border controls may encourage cross-border employment by decreasing travel costs in terms of average and the variance of commuting time. The Freedom of Movement abolished the requirement of a work permit within the European Union (EU) as well as the requirement to prioritize local job seekers over other Europeans. This policy is expected to encourage job search across the border and widen job market opportunities of the poorer areas. 

Our identification strategy relies on the staggered entry in the Schengen and the Freedom of Movement of European countries. Using European Labour Force Survey (ELFS) data aggregated at the regional-level for the period 2005-2019, we estimate their effect on the share of cross-border workers in sending border regions. We identify treatment effects by comparing sending regions to other border regions that maintained a stable status during the observational period. As most of the available control regions in the ELFS have already entered Schengen and the Freedom of Movement in 2005 when we start the analysis, we focus on these border regions, which make our control group composed of always-treated units.  We provide evidence that opening the border almost doubled the probability of working beyond the border in Eastern European regions. The finding is consistent with previous studies that find opening the border increases the share of non-resident workers in host countries \citep[e.g.,][]{Dustmann16,Beerli20}.

The paper is structured as follows. We first provide background on the Schengen and the Freedom of Movement Agreements in Section \ref{institution}. Sections \ref{identification} introduces the setup of the retrospective causal estimation problem and the general matrix completion method. Section \ref{estimation} discusses the specific modeling and estimation procedure adapted to the empirical application. Section \ref{sec:application} describes the data and presents the results for the empirical application. Section \ref{conclusion} concludes. 

\section{Institutions} \label{institution}

Since its foundation, the European Union (EU) has been based on the integration of different economic systems through the removal of obstacles to the movement of people, goods, capital, and services. Geographical mobility of people has been a precondition to the creation of the European common currency, the Euro, because an optimal monetary area requires free movement of inputs to absorb regional shocks \citep[e.g.,][]{mundell1961theory}, smooth spatial mismatches \citep[e.g.,][]{Kain68, Gobillon07}, and attenuate differences in the regional labour markets \citep[e.g.,][]{borjas2001does,niebuhr2012does}.  

In this section, we firstly describe the institutional formation and contemporary status of two milestones of the European integration, the Schengen Agreement and the Freedom of Movement (FoM). Secondly, we mention the findings of the previous empirical literature on the effect of these institutions.

\subsection{Schengen Agreement} \label{Schengen}

The Schengen Agreement was signed in 1985 by the Benelux countries, France and West Germany. This agreement had the goal of gradually abolishing the passport controls, harmonizing the visa policy and enhancing security coordination between countries. The crossing of the border was facilitated, and border controls were replaced by visual checks on private vehicles at a reduced speed. The full abolition of the border controls in the internal EU border and the creation of a single external border came into force in 1995, when also Spain and Portugal joined the Area.  To access the Schengen Area, the countries had to meet specific requirements in the terms of external border controls, issuing visas, police cooperation and personal data protection. 

The Schengen Agreement operated independently from the EU legislation until 1999 when it was incorporated in the Treaty of Amsterdam. Most of the EU countries progressively joined the Schengen Area, while four EU countries, namely Bulgaria, Croatia, Cyprus, and Romania, are not yet part of it. Today, the Schengen Area covers about 425 million people and 4.3 million square kilometre area and includes twenty-two EU countries and four European Free Trade Association (EFTA) members, including Iceland, Liechtenstein, Norway, and Switzerland.

\subsection{Freedom of Movement (FoM)} \label{freedom}

The Freedom of Movement (FoM) in Europe abolished the requirement of a work permit within the EU as well as the requirement to prioritize local job seekers over other Europeans. Differently from the Schengen Agreement, the FoM was always an integral part of the EU integration and therefore applied automatically to all member states. EU enlargements were usually followed by temporary restrictions from some countries such as the ten eastern European countries entering the EU in 2004 (the so-called A10).  While most of the A10 countries granted access to their labour market between 2004 and 2007,  Austria and Germany, which shared the border with five A10 countries, maintained their system of quotas and work permits until 2011, which we exploit in our identification strategy. Switzerland, a non-EU country, progressively granted labour market access to EU residents.  In 2004, full labour market access was granted to cross-border workers working in the regions of Switzerland close to the border. In 2007, this was extended to all regions of Switzerland and the FoM was granted also to EU migrants, with the abolition of the annual quotas.  

\subsection{Related literature} 

According to the economic theory, European institutions such as the Schengen Agreement and the FoM may affect cross-border commuting. By decreasing travel costs in terms of average and the variance of commuting time, the Schengen Agreement may encourage crossing of borders. Similarly, abolishing the requirement of a work permit as well as the requirement to prioritize local job seekers over other Europeans, the FoM may widen job marker opportunities of individuals living in poorer areas beyond the border.

However, studies on the effects of these institutions on cross-border employment of sending regions are scarce.  Most of the literature has focused on their effects on migration \citep[e.g.,][]{elsner2013does,beine2019aggregate}, trade integration \citep{CHEN2011206, Davis14, Felbermayr18},  cross-regional labour market correlations \citep{bartz2012role}, labour market outcomes of natives in the host regions \citep{naguib2019, Basten19, aepli2019,  Aslund19, Beerli20}, and criminality \citep{Sandner18}.

An exception in the literature is \citet{parentischenghen}, who assess the effect of the Schengen Agreement on the cross-border incidence in the nearby sending regions. The authors rely on ELFS data and restrict their sample on individuals working in another region than the one of residence. They implement a  standard difference-in-differences estimator by comparing the incidence of cross-border employment out of total commuters in Italian, French and German regions who shared a border with Switzerland to a control group of regions from the same countries which instead shared the border with other countries that had already accessed Schengen at the beginning of the analysis (always-treated). The authors find an increase in the probability of cross-border employment (conditional on commuting) following the entry of Switzerland in the Schengen Area, while finding no effect attributable to the FoM.

The empirical literature has also shown that other socio-economic factors  may affect commuting across the border. \citet{Wintr09} state that in the Greater Region of Luxembourg the border discourages commuting flows due to language differences. Cross-border commuting is instead more likely to occur the wider the asymmetries in the local labour markets and the easier the accessibility.  Similar results are also found by \citet{AHRENS2020} on the Irish border, which suggests that socio-cultural factors may discourage cross-border commuting despite the absence of custom and border checks. This is also found for commuting across internal regions in Belgium, which confirms that language and cultural differences can discourage commuting \citep{Persyn15}. \citet{Egger15} show that internal language borders in Switzerland can also affect trade. Finally, \citet{Bello19} shows that in Switzerland variations in the exchange rate may also influence the extensive and intensive margin of cross-border labour supply.

\subsection{Staggered implementation}\label{stag}

To assess the effects of these European institutions, we rely on their staggered implementation for the different regions of the EU. We focus on two clusters of sending regions  which entered these institutions between 2007 and 2011 --- namely, the regions of Eastern Europe and Switzerland. The first group is composed of external regions in  Poland, Czech Republic, Slovakia, Slovenia and Hungary, which share a border with Germany or Austria and accessed Schengen in the 1st quarter of 2008 and the FoM with Germany and Austria in the first quarter of 2011. Second, the regional cluster around Switzerland (Italy, France, Germany and Austria) fully entered the FoM in the second quarter of 2007 and accessed Schengen in the first quarter of 2009.

To identify treatments effects, we compare these two sets of regions (later-treated) to the other border regions that maintained a stable status during the observational period. Differently from the previous literature, in our evaluation setting there is virtually no never-treated unit since most of the regions have already entered the treatments at the beginning of the analysis. The regions of Romania and Bulgaria could compose the never-treated group for the Schengen Agreement. However, their treatment status is also not stable since they accessed the EU (and the FoM) during the analysis. Furthermore, their cross-border flow to bordering countries is close to zero. Therefore, we focus on these border regions, which make our control group composed of always-treated units. This identification strategy is formalized in the following section.

\section{Retrospective causal estimation and the general matrix completion strategy}\label{identification}

In our empirical application, we have later-treated units --- i.e., Eastern cluster and Swiss border regions --- and always-treated units --- i.e., regions that maintained a stable treatment status throughout the study period --- and we are interested in estimating the effect of a policy intervention (i.e., European integration) on the later-treated units.

To formalize our setting, consider a sample of $N$ units, each of which was observed in $T$ periods of time. The units are divided into two groups: the always-treated (AT) group, where $N_{\text{AT}}$ are treated throughout the study period, and the later-treated (LT) group, where $N_{\text{LT}}$ units are exposed to the treatment at the initial treatment period, $T_{0} (<T)$; and $N=N_{\text{AT}}+N_{\text{LT}}$. For each unit $i$ and at each time $t$, we observe an outcome $Y_{it}$, a set of covariates $X_{it}$, and the treatment indicator $W_{it}$. For the always-treated units, $W_{it}=1$ for all $t$; for the later-treated units, $W_{it}=0$ for $t < T_0$ (denoted by ``pre'' period) and $W_{it}=1$ for $T_0 \leq t\leq T$ (denoted by ``post'' period). 

Under the Neyman-Rubin potential outcomes framework \citep{neyman1923,rubin1974estimating}, for each unit $i$ and time $t$, there exists a pair of potential outcomes, $Y(1)_{it}$ and $Y(0)_{it}$, corresponding to potential outcomes under the treatment and control condition, respectively. The potential outcomes framework
implicitly assumes treatment is well-defined to ensure that each unit has the same number of potential outcomes. It also excludes interference between units, which would undermine the framework by creating more than two potential outcomes per unit, depending on the treatment status of other units. The fundamental problem is that we can only observe $Y(1)_{it}$ or $Y(0)_{it}$ for each $(i,t)$ pair of indices.

In the retrospective setting, the causal estimand is the average treatment effect for the later-treated units and its average: 
\begin{equation} \label{eq:ATT}
	\tau_{t}^{LT, pre}=E[Y(1)_{it}-Y(0)_{it}], \quad \forall \, i \in LT, t <T_0; \qquad \text{and} \qquad   \tau^{LT, pre}= E_t[\tau_{t}^{LT, pre}].
\end{equation}
In the above estimand, the potential outcome under control in the pre-treatment period is observed, but the counterfactual outcome of the later-treated units under treatment in the pre-treatment period is unobserved and has to be estimated from the observed quantities. In our empirical application, the estimand can be interpreted as the causal effect of European integration on the later-treated regions in the years prior to the implementation of Schengen and FoM, as if they had been always-treated.  

We now introduce the matrix notation of the above setup. Let $\mathbf{Y} \in \mathbb{R}^{N \times T}$ be the matrix of outcomes with typical element $Y_{it}$, and $\mathbf{W} \in \mathbb{R}^{N \times T}$ be the treatment assignment matrix with typical element $W_{it} \in \{0,1\}$. In the retrospective setting, we partition $\boldsymbol{Y}(1)$, the $N \times T$ matrix of the potential outcome under treatment, by treatment group and periods. In order to identify the causal estimand, $\tau_{t}^{LT, pre}$, we need to impute the missing quadrant using information from the observed part of the matrix:
\begin{equation}
	\boldsymbol{Y}(1) =
	\left(\begin{array}{c|c}
		\boldsymbol{Y}(1)_{\text{AT, pre}} & \boldsymbol{Y}(1)_{\text {AT, post}} \\
		\hline	? & \boldsymbol{Y}(1)_{\text {LT, post}}
	\end{array}\right), \label{eq:Y_decomp}
\end{equation}
where the question mark corresponds to the missing potential outcomes, $\boldsymbol{Y}(1)_{\text {LT, pre}}$. 

We illustrate the idea of retrospective causal estimation by exploiting the similarity between the matrix completion and interactive fixed effects models, which are both specified in terms of time-varying factors and unit-specific factor loadings. Following the steps of the interactive fixed effects estimation procedure is effective to clarify how the retrospective analysis works. We first write the potential outcomes under treatment $\boldsymbol{Y}(1)$ as:
\begin{align}
	Y(1)_{it}&=\sum_{r=1}^{R} \gamma_{ir} \delta_{tr}+\epsilon_{it}, \nonumber \\ 
	\qquad \text{or } \mathbf{Y}(1)&= \mathbf{U} \mathbf{V}^{\top}+\boldsymbol{\epsilon}, 
\end{align}
where $\mathbf{U}_{N \times R}$ is a matrix of factor loadings (i.e., unit-specific intercepts) with typical element $\gamma_{ir}$, $\mathbf{V}_{T \times R}$ is a matrix of factors (i.e., time-varying coefficients) $\delta_{tr}$, and $\boldsymbol{\epsilon}_{N \times T}$ is a matrix of random noise $\epsilon_{it}$. In the literature on interactive fixed effects models \citep[e.g.,][]{bai2003inferential,bai2009panel,gobillon2016regional,xu2017generalized}, estimating $\mathbf{U}$ and $\mathbf{V}$ relies fixing the number of factors, $R$. In the matrix completion literature \citep[e.g.,][]{candes2009exact, mazumder2010spectral, recht2011simpler}, it is only assumed $\mathbf{U} \mathbf{V}^{\top}$ is low rank. 

In the retrospective setting, we partition $\mathbf{U}$ and $\mathbf{V}$ analogous to Eq. \eqref{eq:Y_decomp}, 
\begin{equation}
	\mathbf{U}\mathbf{V}^{\top}=\left(\begin{array}{l}
		\mathbf{U}_{\text{AT}} \\
		\mathbf{U}_{\text{LT}}
	\end{array}\right)\left(\begin{array}{l}
		\mathbf{V}_{\text{pre}}^{\top} \\
		\mathbf{V}_{\text{post}}^{\top}
	\end{array}\right).
\end{equation} 
Using data from the always-treated in both periods, and post-treatment data for the later-treated, we can then model the observed part of the $\bY(1)$ matrix as:
\begin{align}\label{IFEsY}
	\mathbf{Y}(1)_{\text{AT}}&=\mathbf{U}_{\text{AT}}\left(\begin{array}{l}\mathbf{V}_{\text{pre}}^{\top} \\ \mathbf{V}_{\text{post}}^{\top}\end{array}\right)+\boldsymbol{\epsilon}_{\text{AT}},
	\qquad \text{and} \qquad \mathbf{Y}(1)_{\text{LT, post}}&=\mathbf{U}_{\text{LT}} \mathbf{V}_{\text{post}}^{\top}+\boldsymbol{\epsilon}_{\text{LT, post}}.
\end{align}

\citet{xu2017generalized} proposes a two-step method, where the left-hand-side equation in \eqref{IFEsY} is an estimator for $\boldsymbol{U}_{\text{AT}}$, $\boldsymbol{V}_{\text{pre}}$, and $\boldsymbol{V}_{\text{post}}$, and the right-hand-side equation is an estimator for $\boldsymbol{U}_{\text{LT}}$ given $\hat{\mathbf{V}}_{\text {post }}$, both by regularized least squares,
\begin{align}\label{IFEs-est}
	\min _{\mathbf{U}_{AT}, \mathbf{V}_{\text {pre }}, \mathbf{V}_{\text {post }}} \left\| \mathbf{Y}_{AT}-\mathbf{U}_{AT}\left(\begin{array}{c}
		\mathbf{V}_{\text {pre }}^{\top} \\
		\mathbf{V}_{\text {post }}^{\top}
	\end{array}\right) \right\|_0
	\qquad \text{and} \qquad \min _{\mathbf{U}_{LT}}\left\|\mathbf{Y}_{LT, \text {post}}-\mathbf{U}_{LT} \hat{\mathbf{V}}_{\text {post }}\right\|_0,
\end{align}
where $\norm{\cdot}_0 = \sum_{i} \mathbf{1}_{\sigma_{i}(\cdot)>0}$ is the rank norm. \citet{athey2017matrix} note that the left-hand-side equation in \eqref{IFEs-est} is not an efficient estimator for $\mathbf{V}$ because it only uses information from the always-treated and does not use $\boldsymbol{Y}_{\text {LT, post}}(1)$ when estimating unobserved factors. Solving Eq.~\eqref{IFEs-est} yields $\hat{\mathbf{U}}_{\text{LT}}$ and $\hat{\mathbf{V}}_{\text{pre}}$, which is used to predict the missing potential outcomes:
\begin{equation}\label{eq:Yhat}
	\hat{\mathbf{Y}}(1)_{\text{LT, pre}}=\hat{\mathbf{U}}_{\text{LT}} \hat{\mathbf{V}}_{\text{pre}}^{\top},
\end{equation}
making feasible to estimate, $\hat{\tau}_{t}^{LT, pre} = \hat{\mathbf{Y}}(1)_{\text{LT, pre}}-\mathbf{Y}(0)_{\text{LT, pre}}$. The estimate $\hat{\mathbf{Y}}(1)_{\text{LT, pre}}$ can be thought as a \emph{retrospective imputation}.

\section{Estimation} \label{estimation}

In this section, we describe our estimation strategy of estimating retrospective causal effects and extensions of the matrix completion method proposed by \citet{athey2017matrix} to accommodate the retrospective setting. 

\subsection{Outcome model}

In the empirical application, the outcome of interest is the share of residents working in another country that shares the border with the region of residence, conditional or unconditional on employment. We model the share of cross-border workers under treatment as:
\begin{equation}
	Y(1)_{it} = L_{it} + X_{it} \beta_t + \gamma_{i} + \delta_{t} + \epsilon_{it}, \label{eq:mc-Y}
\end{equation} 
where $\mathbf{L}$ is an unknown matrix to be estimated, with typical element $L_{it}$. Eq.~\eqref{eq:mc-Y} is identical to the outcome model of \citet{athey2017matrix}, except we allow for a unit- and time-varying covariate matrix $\mathbf{X}$ with typical element $X_{it}$ and estimate a $T$-length vector of time-varying covariate coefficients, $\beta_t$. In the application, $\mathbf{X}$ takes the form of regional GDP divided by the highest GDP of the nearby foreign regions. Unit-specific fixed effects, $\boldsymbol{\gamma}=\{\gamma_1,...,\gamma_N\}$, and time-specific fixed effects, $\boldsymbol{\delta}=\{\delta_1,...,\delta_T\}$ control for unit-specific (time-invariant) and time-specific (unit-invariant) unmeasured confounding. The identifying assumption is that the errors $\epsilon_{it}$ are independent across $i$, are conditional mean zero, and are conditionally independent on $W_{it}$ \citep{xu2017generalized}:
\begin{equation}
	\bE(\epsilon_{it} | L_{it}, X_{it}, \gamma_{i}, \delta_{t}) = \bE(\epsilon_{it} | W_{it}, L_{it}, X_{it}, \gamma_{i}, \delta_{t}) = 0. \label{eq:exog_ass}
\end{equation} 
This exogeneity assumption is analogous to the unconfoundedness assumption in the literature on semiparametric difference-in-differences estimators \citep{abadie2005semiparametric}; namely, that unconfoundedness is conditional on covariates and fixed effects, the latter of which are imposed to capture unmeasured confounding \citep{angrist2009}.

The estimation problem is to recover the entire matrix $\mathbf{L}$ despite the missing part of the matrix, $\boldsymbol{Y}_{\text {LT, post}}(1)$. It is assumed that $\mathbf{L}$ can be approximated by a low-rank matrix, $\mathbf{L}_{N \times T} \approx \mathbf{U}_{N \times R} \mathbf{V}^\top_{T \times R}$, where $R$ is the number of non-zero singular values $\sigma_i (\mathbf{L})$ recovered from a singular value decomposition of the matrix, $\hat{\boldsymbol{L}}_{N \times T} = \boldsymbol{S}_{\mathrm{N} \times \mathrm{N}} \boldsymbol{\Sigma}_{\mathrm{N} \times \mathrm{T}} \boldsymbol{R}^{\top}_{\mathrm{T} \times \mathrm{T}}$, where $\boldsymbol{S}$ and $\boldsymbol{R}$ are unitary matrices and $\boldsymbol{\Sigma}$ is a rectangular diagonal matrix with $\sigma_i (\mathbf{L})$ on the diagonal. 

Let $\mathcal{O}$ denote the set of observed values; i.e., the values for which $W_{it} =1$. \citet{athey2017matrix} propose a nuclear norm estimator for $\mathbf{L}$ in the standard ``forward-looking'' approach that we reformulate below for the retrospective analysis:
\begin{equation}
	\argmin_{\mathbf{L}, \boldsymbol{\beta}, \boldsymbol{\gamma}, \boldsymbol{\delta}} \Bigg[\frac{1}{|\mathcal{O}|} \sum_{(i,t) \in \mathcal{O}} \, \frac{1-\hat{w}_{it}}{\hat{w}_{it}}  \, \bigg(Y_{it}(1) - L_{it} - X_{it} \beta_t - \gamma_{i} - \delta_{t} \bigg)^2
	+ \lambda_L \norm{\mathbf{L}}_\star + \lambda_\beta \norm{\boldsymbol{\beta}}_1 \Bigg], \label{eq:mc-opt-prop}
\end{equation}
where $\hat{w}_{it}$ is the individual predicted probability to be treated at time $t$, and the nuclear norm, or sum of singular values, $\norm{\cdot}_\star = \sum_{i} \sigma_i (\cdot)$, is used to yield a low-rank solution for $\mathbf{L}$ by decomposing $\mathbf{Y}(1)$ in factors and factor loadings. The vector $\ell_1$ norm, $\norm{\cdot}_1$, is used to prevent overfitting by shrinking the values of $\beta_t$ toward zero. The values of the penalty hyperparameters $\lambda_L$ and $\lambda_\beta$ are chosen by cross-validation.

The exogeneity assumption (\ref{eq:exog_ass}) and the goal to impute the missing potential outcomes, $\boldsymbol{Y}(1)$, $(i,t) \in \Bar{\mathcal{O}}$, creates an analogy with a weighting estimation procedure under unconfoundedness \citep{Hirano2001,li2018balancing}. Specifically, the weights $\frac{1-\hat{w}_{it}}{\hat{w}_{it}}$ in (\ref{eq:mc-opt-prop}) balance the observed covariate and unobserved fixed effects distributions between different groups when the target population is the untreated units. These weights place more emphasis on the loss, in (\ref{eq:mc-opt-prop}), for the unit-time elements $(i,t) \in \mathcal{O}$ most similar to the elements $(i,t) \in \Bar{\mathcal{O}}$ in terms of both observed covariates and unobserved fixed effects.

To quantify $\hat{w}_{it}$,  we model treatment as a linear function of a low-rank representation of treatment, $\mathbf{E}$, covariates $X_{it}$, time-varying coefficents $\phi_t$, and fixed effects $\xi_{i}$ and $\psi_{t}$:
\begin{equation}
	W_{it} = E_{it} + [X_{it}  I(W_{it}=1) + \hat{X}_{it} I(W_{it} =0)] \phi_t + \xi_{i} + \psi_{t} + \upsilon_{it}, \label{eq:mc-e-w} 
\end{equation}
where $\hat{X}_{it}$ is an imputed value for ${X}_{it}$ to correct for endogeneity of covariates, as detailed in the following subsection, and $I (\cdot)$ denotes the indicator function. This expression is similar to the treatment model proposed by \citet{athey2017matrix}.

We then estimate the matrix of the treatment assignments, $\mathbf{E}$, analogously to $\mathbf{L}$ in the outcome model \eqref{eq:mc-Y}, by nuclear norm regularized least squares:
\begin{equation}
	\begin{split}
		\argmin_{\mathbf{E}, \boldsymbol{\phi}, \boldsymbol{\xi}, \boldsymbol{\psi}} \Bigg[\sum_{(i,t)} \frac{1}{NT} \bigg(W_{it} - E_{it}  - [X_{it}  I(W_{i,t}=1) + \hat{X}_{i,t} I(W_{i,t} =0)] \phi_t - \xi_{i} - \psi_{t} \bigg)^2 +\\
		\lambda_L \norm{\mathbf{E}}_\star + \lambda_\phi \norm{\boldsymbol{\phi}}_1\Bigg], \label{eq:mc-opt-e}
	\end{split}
\end{equation}
This way $\mathbf{E}$ is the product of a matrix containing $R$ unobserved treatment assignment factors (unit-invariant), $\mathbf{H}_{T \times R}$, and a matrix of unit (time-invariant) factor loadings, $\mathbf{\Lambda}_{N \times R}$, so that any unit loads each treatment assignment factors differently. Once $L_{it}, \beta_t, \gamma_{i}, \delta_{t}$ have been estimated, given assumption (\ref{eq:exog_ass}), we can predict $\hat{Y}(1)_{it}$ for the later-treated in the pre-treatment period by
\begin{center}
	$\hat{Y}(1)_{it} = \hat{L}_{it} + \hat{X}_{it} \hat{\beta}_t + \hat{\gamma}_{i} + \hat{\delta}_{t}$,
\end{center}
for $(i,t)$ such that $i \in LT$ and $t < T_0$.

\subsection{Imputing endogenous covariate values} \label{endogenous}

The minimization problem in Eq. \eqref{eq:mc-opt-prop} requires unit covariate values that are observed both at the pre- and post-treatment periods. Including unit characteristics that are possibly associated with the dynamics of the outcome helps to avoid a strict parallel trends assumption, such as in difference-in-differences analyses. On the other hand, the outcome model (\ref{eq:mc-Y}) imposes restrictions on the choice of the covariates. Like in standard difference-in-differences analyses, the model requires the so-called exogeneity condition for the covariates, which imposes that they are not affected by the treatment: $X_{it}=X(0)_{it}=X(1)_{it}$, where $X(w)_{it}$ is the potential quantity for the covariate when $W_{it}=w$, $w=\{0,1\}$  \citep{lechner2010}. However, in situations where the covariates are endogeneous, so that $X(0)_{it} \neq X(1)_{it}$, the observed covariates are $X_{it}=X(1)_{it}  I(W_{i,t}=1) + X(0)_{it}  I(W_{i,t}=0)$, making
\begin{equation}
	Y(1)_{it} = L_{it} + X(1)_{it}  \beta_t + \gamma_{i} + \delta_{t} + \epsilon_{it} \neq L_{it} + X_{it} \beta_t + \gamma_{i} + \delta_{t} + \epsilon_{it},
\end{equation}
for $(i,t)$ such that $i \in LT$ and $t < T_0$.

We deal with the problem of endogenous covariate values by treating endogeneous values of the covariates the same way we consider the missing potential outcomes, namely as intentional missing data, and impute the endogenous covariate values as follows. We perform the matrix completion procedure in Eqs. \eqref{eq:mc-Y} and \eqref{eq:mc-opt-prop}, but without covariates and instead considering the covariate matrix $\mathbf{X}$ as the outcome. We make the values of the covariate for the later-treated before treatment intentionally missing and replace the missing values with the corresponding values, $\hat{X}_{it}$, from the reconstructed matrix.

\subsection{Elapsed-time weighted loss function} \label{weighted}

A possible complication for our retrospective analysis arises from the fact that we observe the outcomes under treatment, $\mathbf{Y}(1)$, for the later-treated and always-treated groups in the same calendar time $t$ but the elapsed time since $T_0$, $z_t$, is different for the two groups. This means that the observed evolution of $\mathbf{Y}(1)$ of the two groups might not be just linked to the different effect of $t$, but also to $z_t$. The latter is particularly important if the treatment effect takes some time before showing its full potential. According to economic theory, it is likely that the treatment effect stabilizes after some periods in a new ``steady state" equilibrium. For example, it is possible that workers take time before adjusting their expectations of the job possibilities in the other countries and spreading the news to the other workers. The influence of $z_t$ may be an issue in the retrospective setting, where we predict $\mathbf{Y}(1)_{LT, \text{pre}}$ by elaborating the information from the three time-series $\mathbf{Y}(1)_{LT, \text{post}}$, $\mathbf{Y}(1)_{AT, \text{post}}$, and $\mathbf{Y}(1)_{AT, \text{pre}}$, if only the two latter series are in a steady state equilibrium.

We address this problem in the retrospective analysis by weighting the objective function so that more weight is placed on the loss for observed values with higher values of $z_t$. More specifically, we use the product of $\hat{w}_{it}$ and $z_t$ to weight the objective function \eqref{eq:mc-opt-prop}. To adjust for elapsed time since the treatment implementation, we compute $\diag(\mathbf{z}) \mathbf{\hat{W}}$, where $\mathbf{z}$ concatenates two sequence of values drawn from a logistic model, the first sequence descending to $T_0$ and the second sequence ascending from $T_0$. This computation places more weight on the ends of $\mathbf{\hat{W}}$ for the later-treated units. The weights used to adjust the loss for later-treated units in Eq. \eqref{eq:mc-opt-prop} is then
$I_{i \in LT} \frac{1-\diag(z_t) \hat{w}_{it}}{\diag(z_t) \hat{w}_{it}}$.

\section{Empirical Application} \label{sec:application}

In this section, we describe the survey data used for the application; we then estimate the effects of Schengen and FoM on the share of residents working in a neighbouring country, conduct placebo tests to assess if the effects are truly attributable to the treatment, and assess the accuracy of the matrix completion estimator.

\subsection{Data} 

We rely on the harmonized quarterly ELFS, which covers a sample of resident households in the different regions of the EU from the 1st quarter of 2005 to the 4th quarter of 2019. The ELFS is the largest household survey in Europe with about 1.7 million interviews per quarter and an effective sampling rate of 0.29\% in 2017. Individuals living in the household are asked questions on their labour status, working time, education, sex, age, nationality and, importantly for our analysis, the region of residence and the country of work.  Information on the location of residence is usually provided at regional level though some countries provide this information only at an aggregated level. Most countries have a rotation  scheme in place, though the anonymised ELFS microdata do  not  contain the information which would allow tracking people across waves. We focus on the working age population and retain all the individuals aged between 20 and 59. As we rely on panel data methods in the econometric analysis, we then collapse the data at the level of the region of residence and follow the evolution of the aggregated outcomes over time. To enhance harmonization, we start following the regions from the first quarter of 2005, when all countries have transited from a yearly survey (usually in Spring) to a continuous quarterly survey, covering  all weeks of the year. We exclude all regions not sharing an inland border with another EU or Schengen country,  countries for which information on the country of work is missing in any year (Portugal, Ireland, Norway, Sweden and Switzerland) and the treated regions that tend to host the cross-border flows (i.e. regions in Austria, Germany and Italy sharing a border with East of Europe).

\begin{figure}
	\centering
	\includegraphics[width=\textwidth]{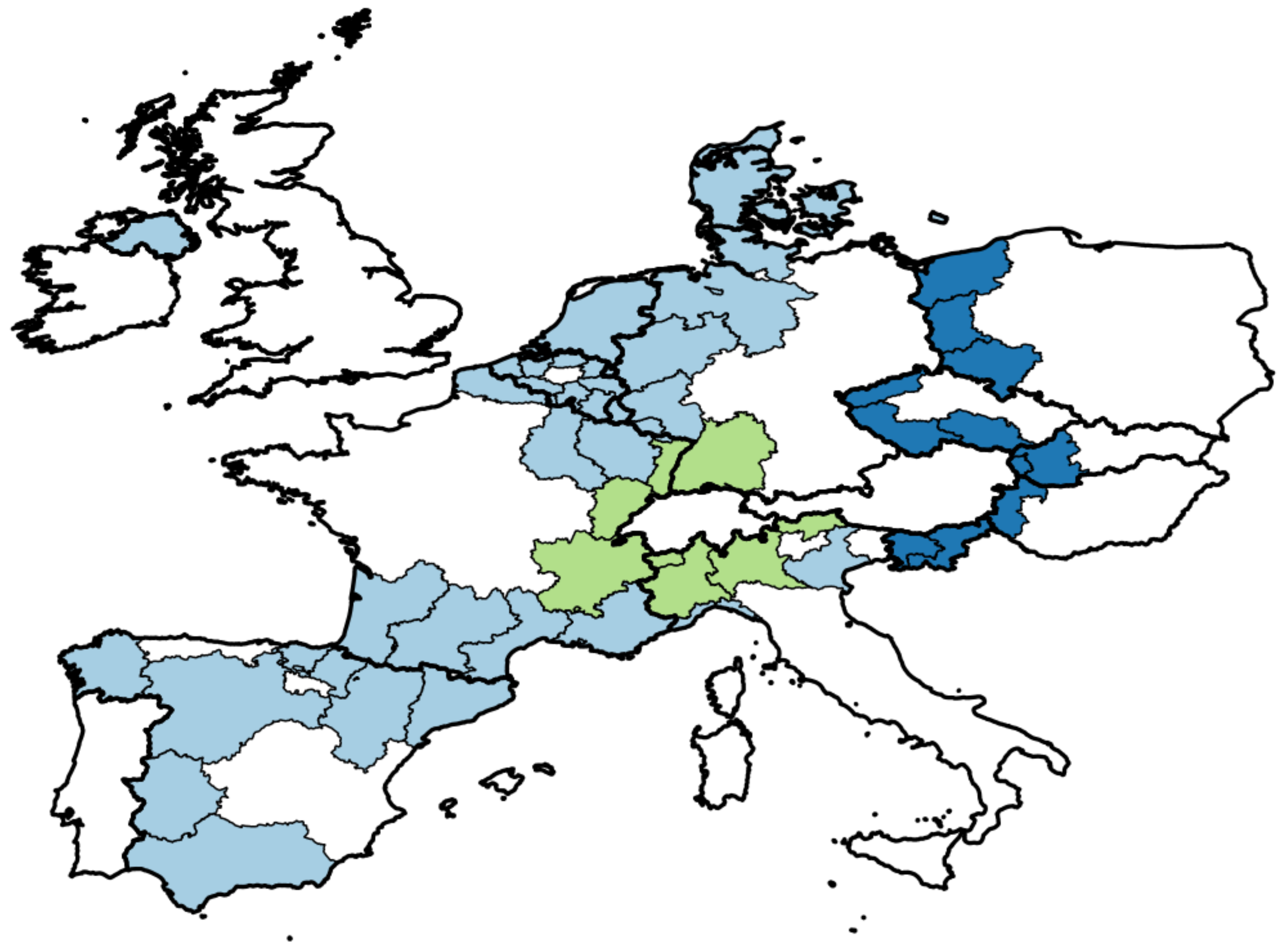}
	\caption{Treated and control regions.  \emph{Key:}	
		{\color{lightblue}{\sampleline{line width=2mm}}}, controls (always-treated); 
		{\color{darkblue}{\sampleline{line width=2mm}}},  Schengen in 2008 and FoM in 2011;
		{\color{lightgreen}{\sampleline{line width=2mm}}},  FoM in 2007 and Schengen in 2009.}\label{treatedCBW}
\end{figure}

The outcome of interest is the share of residents working in another country, which shares the border with the region of residence, conditional or unconditional on employment. As the survey covers only residents, this outcome should include cross-border commuters or temporary migrants who come back to their home country regularly. The treatment statuses and periods, mentioned in Section \ref{stag},  are summarized in Figure \ref{treatedCBW}  and Table \ref{dates}.

\begin{table}
	\caption{\label{dates}Staggered treatments.}	
	\centering
	\begin{tabular}{cccc}
		\toprule
		&  &  &  \\
		Period	&   Eastern cluster  & Swiss cluster & Always-Treated \\
		\midrule
		&  &    &   \\
		2005Q1-2007Q2	& - &  (Mild FoM)  &  Schengen + FoM \\
		2007Q3-2007Q4	& - &  FoM &  Schengen + FoM \\
		2008Q1-2008Q4	& Schengen  &  FoM &  Schengen + FoM  \\
		2009Q1-2010Q4	& Schengen  &   Schengen + FoM &  Schengen + FoM  \\
		2011Q1-2019Q4	& Schengen + FoM &   Schengen + FoM &  Schengen + FoM \\
		\bottomrule
	\end{tabular}
\end{table}

\subsection{Results}\label{results}

We estimate the effect of the European integration on the share of residents working in a neighbouring country. Since there are two treatments in the application --- Schengen and FoM --- we set $T_0$ at the time when both treatments are in place. In the Eastern cluster, the Schengen policy and the FoM  are adopted in the first quarters of 2008 and in 2011, respectively, so we set $T^{(\text{Eastern})}_0 =2011$. In the Swiss cluster, the FoM is adopted in the second quarter of 2007 and the Schengen policy is adopted in the first quarter of 2009, so we set $T^{(\text{Swiss})}_0 = 2009$. 

\begin{figure}
	\centering
	\includegraphics[width=\textwidth]{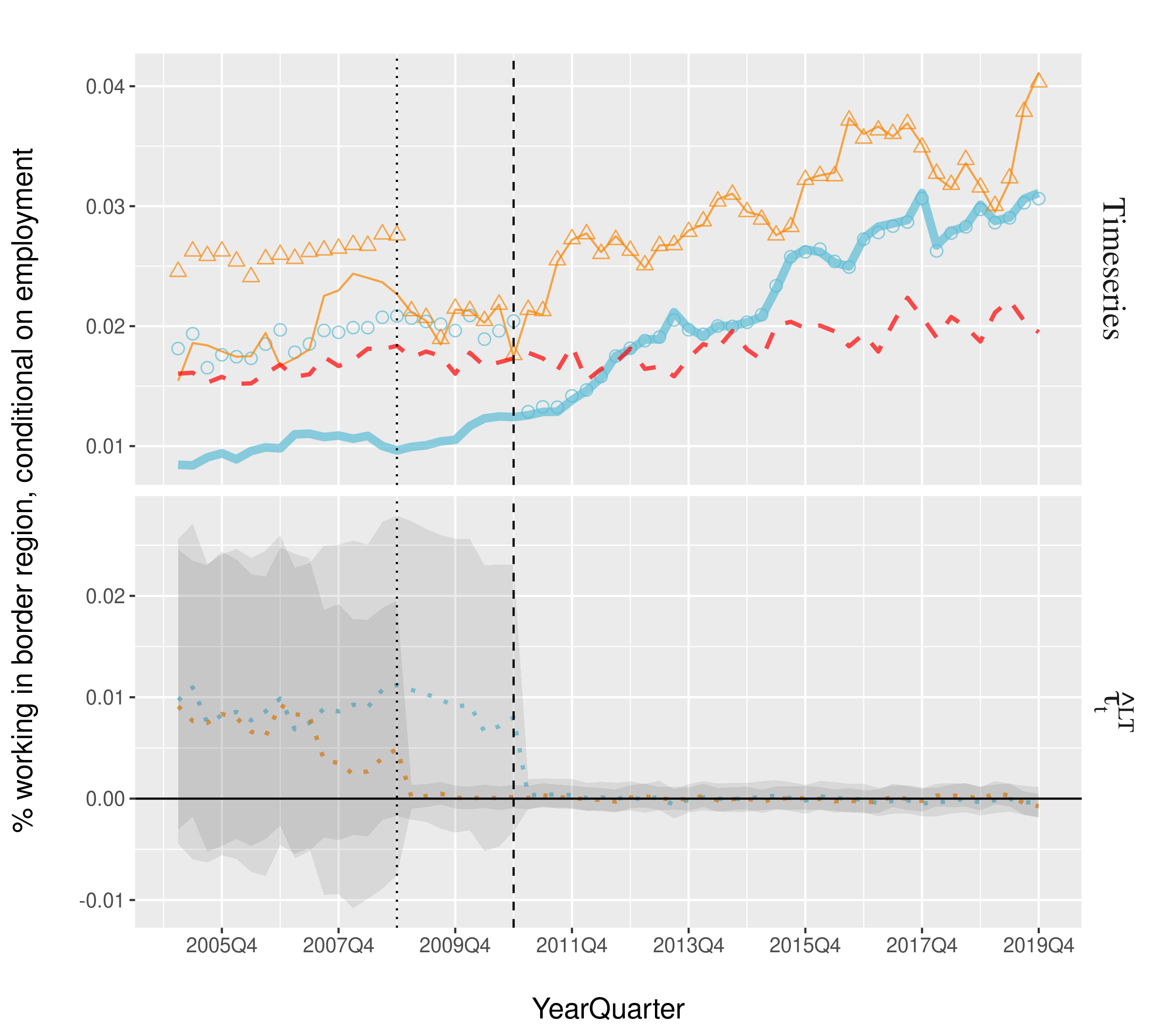}
	\caption{Matrix completion estimates of the effect of combined treatment (Schengen + FoM) on share of residents working in another country, conditional on employment, for the later-treated regions ($N \times T = 53 \times 60$). \emph{Note:} The dashed vertical lines represent the initial treatment times of the combined treatment, $T^{(\text{Eastern})}_0 = 2011$ and $T^{(\text{Swiss})}_0 = 2009$. The shaded regions are 95\% bootstrap confidence intervals of the effects constructed with 999 block bootstrap samples. \emph{Key:}		{\color{Darjeeling15}{\sampleline{line width=1mm}}}, observed outcomes of Eastern regions;
		{\color{Darjeeling14}{\sampleline{line width=0.25mm}}}, observed outcomes of Swiss regions;
		{\color{Darjeeling11}{\sampleline{dash pattern=on .6em off .4em on .6em}}}, observed outcomes of always-treated regions;
		{\protect\tikz \protect\draw[color={rgb:red,91;green,188;blue,214}] (0,0) -- plot[mark=o, mark options={scale=2.5,fill=white}] (0,0) -- (0,0);}, counterfactual Eastern; 
		{\protect\tikz \protect\draw[color={rgb:red,249;green,132;blue,0}] (0,0) -- plot[mark=triangle, mark options={scale=2.5,fill=white}] (0,0) -- (0,0);}, counterfactual Swiss; 
		{\color{Darjeeling15}{\sampleline{dash pattern=on .15em off .15em on .15em off .15em}}}, $\hat{\tau}_t^{\text{Eastern}}$;
		{\color{Darjeeling14}{\sampleline{dash pattern=on .15em off .15em on .15em off .15em}}}, $\hat{\tau}_t^{\text{Swiss}}$. \label{mc-estimates-cbw-CBWbordEMPL}} 
\end{figure}

The top panel of Figure \ref{mc-estimates-cbw-CBWbordEMPL} shows the evolution of the (conditional) cross-border observed outcomes (solid lines) for the two clusters of later-treated regions and the always-treated border regions (dashed line). The descriptive evidence shows an increase in the outcome of both later-treated clusters of regions after the treatments. At the same time, the outcome of the always-treated regions remains quite stable and increases only marginally throughout the  period of analysis. 

We then check whether this suggestive descriptive evidence on a positive effect of the European institutions is confirmed by our matrix completion method. As described, we specify a propensity-score and elapsed-time weighted loss function \eqref{eq:mc-Y}-\eqref{eq:mc-opt-prop}. We then predict the counterfactual outcomes of the later-treated regions, $\mathbf{\hat{Y}}(1)_{\text{LT, pre}}$, using information from the observed outcomes of the later-treated and always-treated. In the bottom panel we plot the estimated treatment effects on later-treated units, $\hat{\tau}_{t}^{LT, pre}$, which are calculated by differencing the predicted outcomes of the later-treated units (dotted lines) and their observed outcomes. The estimated treatment effects are surrounded by percentile bootstrap confidence intervals constructed by block resampling the columns (i.e, time dimension) of the observed outcomes, in order to preserve temporal dependence structure of the original data \citep{davison1997bootstrap,politis2004automatic}. To infer the overall effect of treatment, we estimate percentile bootstrap confidence intervals for the counterfactual trajectory, $\hat{\tau}^{LT, pre}$, by resampling the trajectories without the time component.

The decomposition of the low-rank matrix $\hat{\boldsymbol{L}}_{N \times T}=\hat{\boldsymbol{S}}_{\mathrm{N} \times \mathrm{N}} \hat{\boldsymbol{\Sigma}}_{\mathrm{N} \times \mathrm{T}} \hat{\boldsymbol{R}}^{\top}_{\mathrm{T} \times \mathrm{T}}=\hat{\mathbf{U}}_{N \times R} \hat{\mathbf{V}}^\top_{T \times R}$ helps to explain the imputed outcome trends. The dominant (first) detected factor in $\hat{\mathbf{V}}_{R \times T}$, is mainly loaded by the later-treated (by means of the loading factors in the sub-matrix $\hat{\mathbf{U}}_{LT \times R}$), and visually resembles the imputed outcome trends of the later-treated in Figure \ref{mc-estimates-cbw-CBWbordEMPL}. The rest of the factors are more stationary and are mainly loaded by the always-treated. The graphs for the detected first and second factors for both the Eastern and Swiss blocks are reported in Figure \ref{mc-trend} in the Supplementary Material (SM).

Our results are in line with the descriptive evidence. The counterfactual trajectory estimates, reported in Table \ref{benchmark-estimates}, show that had the Schengen and FoM both been implemented in the Eastern cluster prior to 2011, the share of cross-border workers out of the total working population would have been 0.8 percentage points [0.2, 1.6] higher. Relative to the observed outcome in the same period, this represents an increase of 86\%. We find a positive effect on the share of cross-border workers in Swiss region of about 0.6 percentage points [-0.2, 2.3], which correspond to a relative increase of about 32\%, although this is not statistically significant at the 5\% level. The effects are marginally smaller if we do not condition the outcome on employment.

We compare the matrix completion estimates with difference-in-differences and synthetic control estimates, also reported in Table \ref{benchmark-estimates}. The difference-in-differences estimator is a regression of outcomes on treatment and unit and time fixed effects \citep{athey2018design}. The synthetic control method is a regression of the pre-treatment outcomes of each treated unit on the control unit outcomes during the same periods, with the restrictions of the original synthetic control method of \citet{abadie2010synthetic}, namely zero intercept and non-negative regression weights that sum to one \citep{doudchenko2016balancing,athey2017matrix}. These restrictions are equivalent to imposing a restriction of linear dependence between the loading factors in the context of matrix completion or interactive fixed effects models. We provide the exact form of these estimators in Section \ref{benchmark-estimators}. 

The difference-in-differences and synthetic control method point estimates are within the confidence intervals of the matrix completion estimates. Note that we express the difference-in-differences as a two-way fixed effects as in \citet{athey2018design} to allow a comparability of results over the entire period 2005Q1-2019Q4. This collapses the temporal effects into a single parameter and thus loses the interpretation in terms of inter-temporal dynamics compared to the matrix completion method. The small difference in the estimated effect between difference-in-differences and matrix completion is due to the fact that the difference in the averaged outcomes in the post-treatment period between the later-treated and always-treated (for both the Eastern and the Swiss block) is similar to the total estimated effect from matrix completion. This occurrence is not indicative of whether the parallel trends assumption holds or not.

\begin{table}
	\caption{\label{benchmark-estimates}Counterfactual trajectory estimates.}
	\fontsize{9}{10}\selectfont
	\centering
	\begin{threeparttable}
		\begin{tabular}{@{}lrrrr@{}}
			\toprule
			&  \multicolumn{2}{c}{Eastern}    & \multicolumn{2}{c}{Swiss}    \\
			&  \multicolumn{2}{c}{(2005Q1-2010Q4)}    & \multicolumn{2}{c}{(2005Q1-2008Q4)}  \\ \midrule
			Estimator & \multicolumn{1}{c}{Uncond.} & \multicolumn{1}{c}{Cond.} & \multicolumn{1}{c}{Uncond.} & \multicolumn{1}{c}{Cond.} \\  
			\midrule \\
			Matrix com.  & \multicolumn{1}{r}{0.006 {[}0.002, 0.012{]}}  &	\multicolumn{1}{r}{0.008 {[}0.002, 0.016{]}} & \multicolumn{1}{r}{0.004 {[}-0.002, 0.016{]}}	&  \multicolumn{1}{r}{0.006 {[}-0.002, 0.023{]}} \\
			Diff.-in-diff.           & \multicolumn{1}{r}{0.008 {[}0.002, 0.015{]}}  &	\multicolumn{1}{r}{0.010 {[}0.003, 0.018{]}} & \multicolumn{1}{r}{0.005 {[}-0.003, 0.020{]}}	&  \multicolumn{1}{r}{0.006 {[}-0.003, 0.025{]}} \\
			Synth. con. & \multicolumn{1}{r}{0.007 {[}0.002, 0.013{]}}  &	\multicolumn{1}{r}{0.009 {[}0.003, 0.016{]}} & \multicolumn{1}{r}{-0.001 {[}-0.007, 0.005{]}}	&  \multicolumn{1}{r}{0.001 {[}-0.008, 0.008{]}} \\
			\bottomrule
		\end{tabular}
		\begin{tablenotes}[flushleft]
			\footnotesize
			\item[] \emph{Notes:} Outcome is \% working in border region, unconditional or conditional on employment. Bracketed values represent 95\% percentile bootstrap confidence intervals for $\hat{\tau}^{LT, pre}$.
		\end{tablenotes}
	\end{threeparttable}
\end{table}

We also investigate the role of the two treatments, Schengen or the FoM, in driving the estimated positive effect. To do that, we divide the pretreatment period into two institutional periods for each regional cluster in which either no treatment or only one of the two treatments was implemented. As the estimated counterfactual represents the outcome under both treatments, the effect on the former provides evidence of the joint impact of the two European institutions, while the effect on the latter represents the impact of the treatment that was not yet implemented, conditional on the other treatment. As shown in Table \ref{mc-estimates-period}, the treatment effect for the Eastern cluster fully exhibits during the period when Schengen was already operative, from 2008Q1 to 2010Q4. This finding suggests that the main driver of the estimated effect is the FoM, an institution that has removed the work permit requirement across European countries. 

Similarly, the Swiss cluster shows a very small treatment effect when the FoM was operative (+0.3 percentage points during the period 2007Q3-2008Q4), which is more than double when no treatment was implemented (+0.7 percentage points before 2007Q2), though it remains statistically insignificant at the 5\% level. Removing border controls does not seem sufficient in boosting cross-border employment, which is in contrast with the results of \citet{parentischenghen} for the Swiss cluster. This study however selected only individuals commuting between regions and ignored that the control group is composed of always-treated. 

Finally, when interpreting the results for the FoM in the Swiss cluster we should keep in mind that our estimates only captures the ``top-up'' effects of the full liberalization of 2007. Indeed, as explained in Section \ref{freedom}, from 2004 cross-border workers were already allowed working in the border regions of Switzerland without a work permit, which according to the previous literature had increased the relative incidence of non-resident employment in the border regions of Switzerland \citep[e.g.,][]{Beerli20}. The 2007 reform extended the FoM to the whole Swiss territory and included also migrants. Our estimates provide evidence that the this extension did not have a significant additional impact.

\subsection{Placebo tests and comparison with alternatives}\label{placebo-tests}

To assess whether the estimated effects are attributable to Schengen and FoM rather than other policy changes during the same period, we conduct placebo tests by conducting a ``no treatment" evaluation; that is, we re-run the analysis on the post-treatment data, when no treatment effect is expected. In the first set of placebo tests, we fit the matrix completion estimator on the post-treatment years and randomize the initial treatment period for the actual treated units in both simultaneous or staggered treatment adoption settings. Letting $\hat{\tau}^{LT, pre}_{\pi}$ denote the average causal effects estimated for each block bootstrap sample $\pi \in \Pi$, we estimate nonparametric bootstrap $p$-values under the null hypothesis, $\text{H}_0: |\hat{\tau}^{LT, pre}| = 0$:
\begin{equation}
	\hat{p} = 1 + \frac{1}{\Pi+1} \sum_{\pi \in \Pi} I \left\{|\hat{\tau}^{LT, pre}_{\pi}| > |\hat{\tau}^{LT, pre}| \right\}. \label{eq:pvalue}
\end{equation}
The estimated $p$-values, reported in Table \ref{placebo-ci}, exceed the standard level of statistical significance except for the Swiss cluster when the initial placebo treatment period is two time periods away from the actual initial treatment period. These results provide indirect evidence that the estimated effects are truly attributable to the Schengen and FoM policies instead of other policy changes or spurious errors during the same period. 

We also leverage the post-treatment period---when the true ``treatment" effect is null---to compare the matrix completion estimator with difference-in-differences and synthetic control method. Specifically, we randomly select $N/2$ treated units and impute their missing values following a given initial placebo treatment period. We evaluate the matrix completion estimator in terms of the average root mean squared error (RMSE), comparing actual versus imputed values, and specify a model without covariates to facilitate comparability with the alternative estimators, which do not rely on covariates.

\begin{figure}
	\centering
	\includegraphics[width=0.49\textwidth]{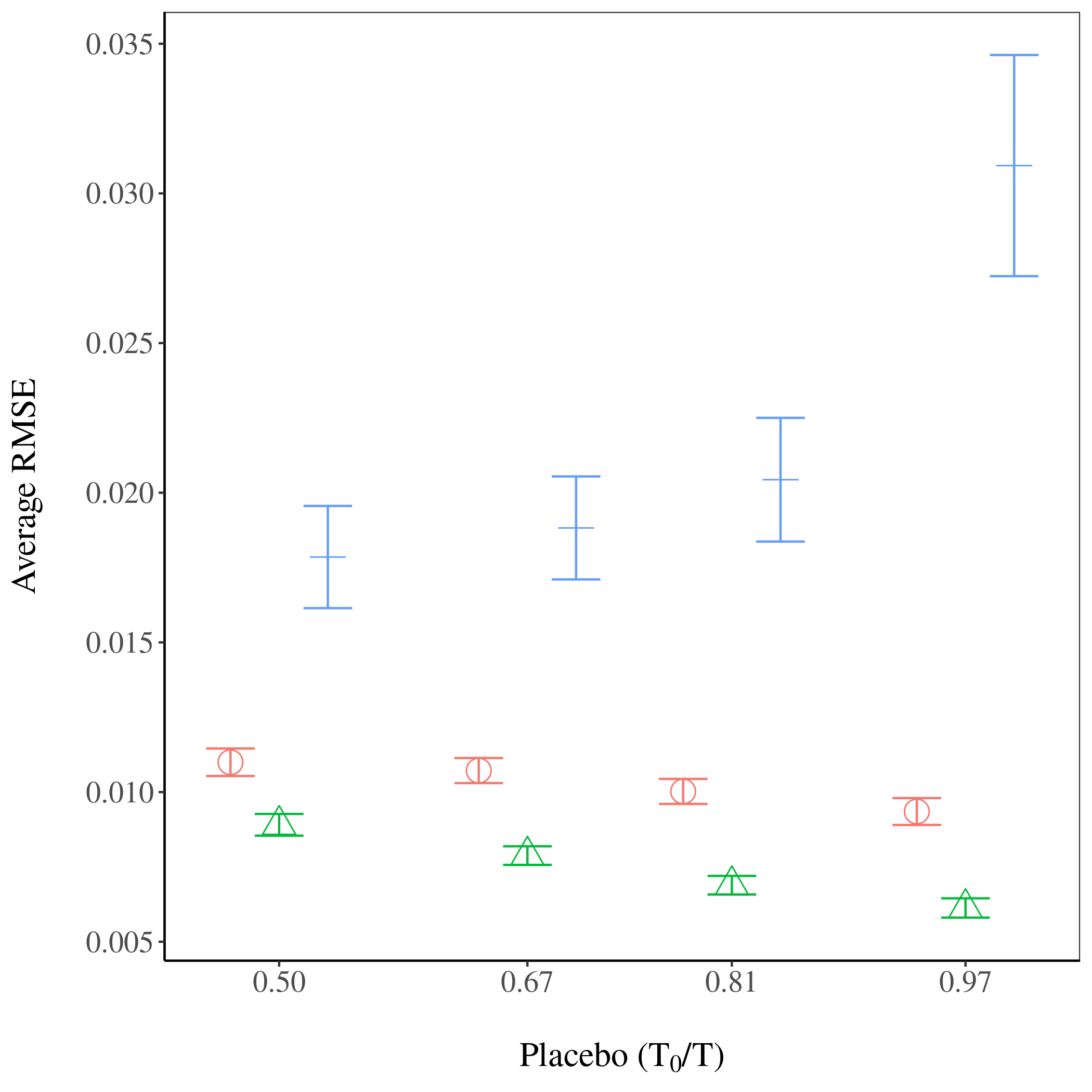}
	\includegraphics[width=0.49\textwidth]{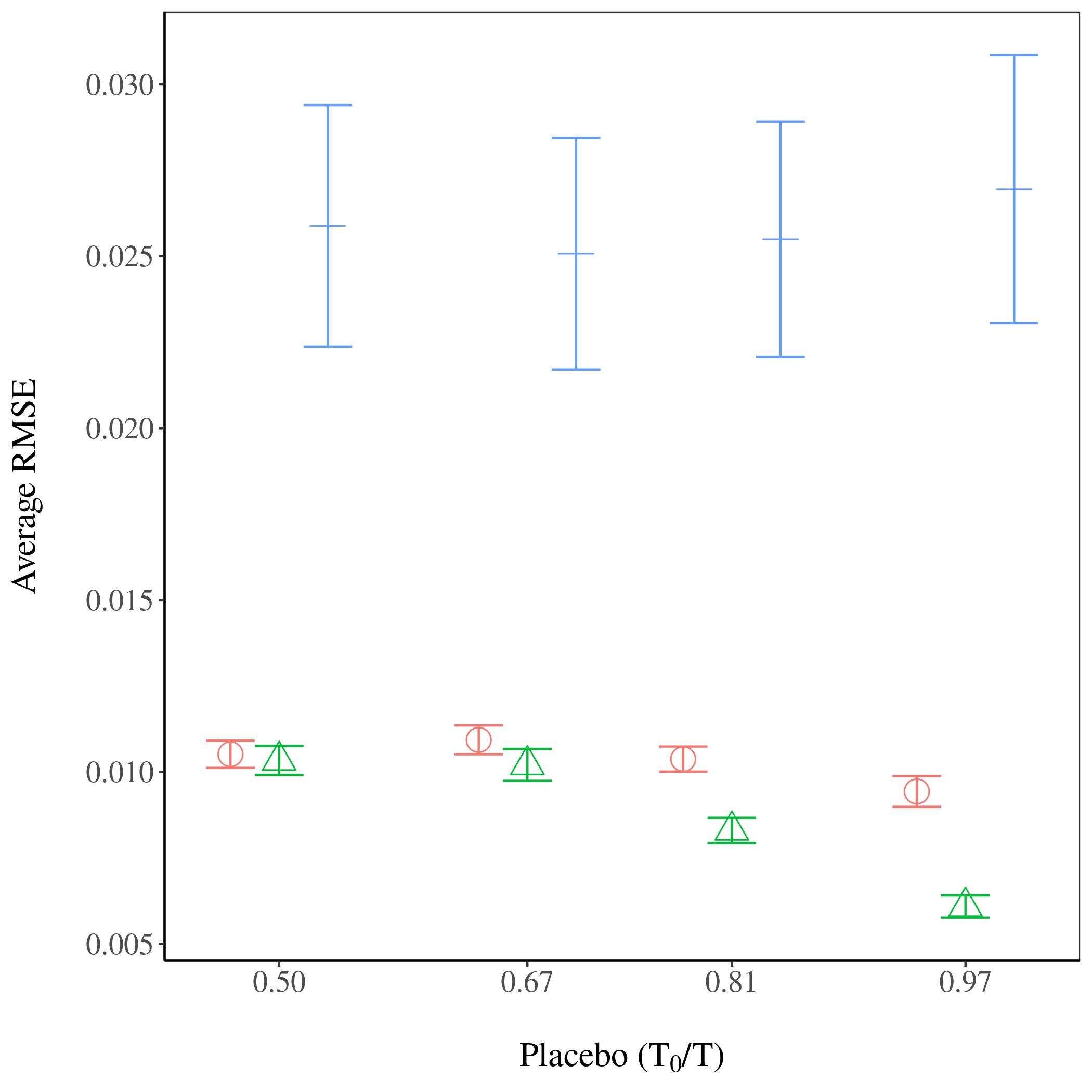} 
	\caption{RMSE for retrospective imputation for placebo treated units on post-treatment data, under staggered treatment adoption (left) or simultaneous treatment adoption (right). Outcome: \% working in border region, conditional on employment. Error bars are calculated by taking average RMSE $\pm$ 1.96 times the standard error of the average RMSE across 100 runs. The $x$ axis is the ratio of the placebo initial treatment time to the number of periods in the placebo data. Estimates are jittered across the $x$ axis to avoid overlap. \emph{Method:} 
		{\protect\tikz \protect\draw[color={rgb:red,0;green,186;blue,56}] (0,0) -- plot[mark=triangle, mark options={scale=3, rotate=0}] (0.25,0) -- (0.5,0);}, Matrix com.;
		{\protect\tikz \protect\draw[color={rgb:red,248;green,118;yellow,109}] (0,0) -- plot[mark=o, mark options={scale=3,fill=white}] (0.25,0) -- (0.5,0);}, Diff.-in-diff.; 
		{\protect\tikz \protect\draw[color={rgb:red,97;green,156;blue,255}] (0,0) -- plot[mark=+, mark options={scale=3, rotate=0}] (0.25,0) -- (0.5,0);}, Synth. con. \label{placebo-rmse-stag}}
\end{figure}

Figure \ref{placebo-rmse-stag} reports the estimators' performance under simultaneous or staggered treatment adoption settings. The $x$ axis is the ratio of the initial placebo treatment time to the number of periods in the placebo data, so higher values represent more training data. In most settings, the matrix completion estimator outperforms difference-in-differences, both of which significantly outperform the synthetic control method. This bolsters the usage of the matrix completion methods in our application.

\section{Conclusion}\label{conclusion}

Motivated by a setting with later-treated and always-treated units, we formulate a novel identification strategy of retrospective counterfactual imputation to predict potential outcomes under treatment for the later-treated in the pre-treatment period. We treat these potential outcomes as missing and impute them using a matrix completion estimator. 

Our methodological contributions are three-fold: first, we introduce a framework for causal inference in the retrospective setting, comparing always-treated and later-treated units; second, we extend the matrix completion estimator by propensity- and elapsed-time weighting the objective function to correct for differences in the observed covariate and unobserved fixed effects distributions, and the elapsed time since treatment between groups. Third, we propose a procedure for imputing endogenous covariate values when reconstructing potential outcomes under treatment.

We apply this framework to evaluate the effect of two key European institutions, namely the Schengen Agreement and the Freedom of Movement, on cross-border employment in Europe. While previous literature has studied the labour market effects of these institutions on natives in the host regions \citep{naguib2019, Basten19, aepli2019,  Aslund19, Beerli20}, our analysis focuses on sending regions.

By relying on ELFS data aggregated at the regional level for the period 2005-2019, we provide evidence that these institutions have almost doubled the probability of working beyond the border for individuals living in the border regions of Eastern Europe. Interestingly, the effect appears to be driven mostly by the Freedom of Movement, which abolished the requirement of work permits, while the abolition of border controls is not the main driver. Our results are consistent with the results of previous studies which showed that opening the border increases the share of non-resident workers in host countries \citep[e.g.,][]{Dustmann16, Beerli20}. The findings are robust to whether difference-in-differences or the synthetic control method is used in place of the matrix completion estimator. Placebo tests provide indirect evidence that the key identification assumption of exogeneity is not violated and that the estimator is consistent.  


\bibliographystyle{rss}
\begin{singlespace}
\begin{footnotesize}
\begin{multicols}{2}
\bibliography{references}
\end{multicols}
\end{footnotesize}
\end{singlespace}

\itemize
\end{document}


\begin{singlespacing}
	\maketitle \thispagestyle{empty}
	\tableofcontents \thispagestyle{empty}
\end{singlespacing}

\setcounter{figure}{0} \renewcommand{\thefigure}{SM-\arabic{figure}}
\setcounter{table}{0} \renewcommand{\thetable}{SM-\arabic{table}}
\setcounter{section}{0} \renewcommand{\thesection}{SM-\arabic{section}}

\pagenumbering{roman}

\pagebreak
\pagenumbering{arabic}

\section{Estimators} \label{benchmark-estimators}

\subsection{DID}

The DID model is specified by \citep{athey2018design}:
\begin{equation}
Y_{it}  = \gamma_i + \delta_t + \tau W_{it} + \epsilon_{it},
\end{equation}
where $\gamma_i$ + $\delta_t$ are unit and time fixed effects, respectively. The model is estimated by least squares: 
\begin{equation}
\argmin _{\tau,\gamma_{i},\delta_{t}} \sum_{i=1}^{N} \sum_{t=1}^{T}\left(Y_{i t}-\gamma_{i}-\delta_{t}-\tau W_{i t}\right)^{2}.
\end{equation}
We use the implementation of DID in the \texttt{MCPanel} \textsf{R} package, available at \url{https://github.com/susanathey/MCPanel}.

\subsection{Matrix completion}

In the placebo tests, the matrix completion model is specified without covariates:
\begin{align}
Y(1)_{it} &= L_{it} + \gamma_{i} + \delta_{t} + \epsilon_{it}, \label{eq:mc-Y-no-covars} \\
\hat{L}_{it} &= \argmin_{\mathbf{L}, \boldsymbol{\gamma}, \boldsymbol{\delta}} \Bigg[\frac{1}{|\mathcal{O}|} \sum_{(i,t) \in \mathcal{O}}  \frac{1-\hat{w}_{it}}{\hat{w}_{it}} \, \bigg(Y_{it} - L_{it} - \gamma_{i} - \delta_{t} \bigg)^2
+ \lambda_L \norm{\mathbf{L}}_\star \Bigg]. \label{eq:mc-opt-prop-no-covars}
\end{align}
We use extend the code of \texttt{MCPanel} to implement the model with the weighted objective function, available at \url{https://github.com/jvpoulos/MCPanel}, under the default settings. In the main analyses, we implement the model with covariates \eqref{eq:mc-Y}-\eqref{eq:mc-opt-prop}, also under default settings. 

\subsection{SCM}

The synthetic control method (SCM) \citep{abadie2003economic,abadie2010synthetic,abadie2015comparative} compares a single treated unit with a synthetic control that combines the outcomes of multiple control units on the basis of their pre-intervention similarity with the treated unit. 

\citet{doudchenko2016balancing} and \citet{athey2017matrix} show that the SCM can be interpreted as regressing the pre-treatment outcomes of a single treated unit on the control unit outcomes during the same periods. The parameters estimated on the controls are used to predict the counterfactual outcomes for a single treated unit, $i=0$:
\begin{align}\label{adh} \nonumber
\hat{Y}_{0t} &= \sum_{i=1}^{\textrm{N}} \hat{\omega}_i Y_{it}, \qquad \forall \, t= \textrm{T}_0, \ldots, \mathrm{T}\\
\textrm{where } \hat{\omega} &=\argmin_{\omega}\sum_{s=1}^{\textrm{T}_0} \left(Y_{0s}-\sum_{i=1}^{\textrm{N}} \omega_i Y_{0s}\right)^2, \qquad \textrm{s.t. } w_i \geq 0, \sum w_i =1.
\end{align}
A separate model is subsequently fit for each $i, \ldots, N_t$ treated units. Note that Eq.~\eqref{adh} imposes the restrictions of the original SCM, namely zero intercept and non-negative regression weights that sum to one. 

We use the \texttt{MCPanel} \textsf{R} package implementation with default settings, except we set the relative tolerance of for the stopping rule to 0.001 and bound gradient values within $[-5,5]$ in order to facilitate convergence. The underlying algorithm for estimating the parameters is exponentiated gradient descent \citep{kivinen1997exponentiated}. 

\section{Tables and Figures} \label{figures-tables}

\begin{table}[htb]
	\centering
	\begin{threeparttable}	\scriptsize
		\caption{Matrix completion estimates by institutional period and cluster.\\ Outcome: \% working in border region, unconditional or conditional on employment. \label{mc-estimates-period}}
		\begin{tabular}{@{}lcccc@{}}
			\toprule
			&  \multicolumn{2}{c}{Eastern}    & \multicolumn{2}{c}{Swiss}    \\
			&  \multicolumn{2}{c}{(2005Q1-2010Q4)}    & \multicolumn{2}{c}{(2005Q1-2008Q4)}  \\ \midrule
			\backslashbox{Period}{Outcome} & \multicolumn{1}{c}{Unconditional} & \multicolumn{1}{c}{Conditional} & \multicolumn{1}{c}{Unconditional} & \multicolumn{1}{c}{Conditional} \\  
			\hline \\
			No Treatment            & \multicolumn{1}{r}{0.006 {[}0.001, 0.012{]}} & 	\multicolumn{1}{r}{0.008 {[}0.002, 0.015{]}}	& \multicolumn{1}{r}{0.005 {[}-0.001,  0.017{]}} & \multicolumn{1}{r}{0.007 {[}-0.001,  0.023{]}} \\
			Single Treatment         & \multicolumn{1}{r}{0.007 {[}0.002, 0.011{]}} &	\multicolumn{1}{r}{0.009 {[}0.003,  0.015{]}}  & \multicolumn{1}{r}{0.002 {[}-0.007,  0.014{]}}  &  \multicolumn{1}{r}{0.003 {[}-0.009,  0.020{]}} \\
			(Eastern: Schengen; Swiss: FoM)      &  &  &		&   \\
			\bottomrule
		\end{tabular}
		\begin{tablenotes}[flushleft]
			\footnotesize
			\item \emph{Notes:}  the no treatment period is 2005Q1-2007Q4 (2005Q1-2007Q2) for the Eastern (Swiss) cluster. The single treatment period is 2008Q1-2010Q4 (2007Q3-2008Q4) for the Eastern  (Swiss) cluster when only Schengen (the FoM) was implemented. Bracketed values represent 95\% percentile bootstrap confidence intervals for $\hat{\tau}^{LT, pre}$.
		\end{tablenotes}
	\end{threeparttable}
\end{table}

\begin{table}[htb]
	\centering
	\begin{threeparttable} \footnotesize
		\caption{Matrix completion placebo test $p$-values. }\label{placebo-ci}
		\begin{tabular}{@{}lcccccc@{}}
			\toprule
			Placebo $T_0/T$                                                                                     & \multicolumn{2}{c}{0.50} & \multicolumn{2}{c}{0.80}& \multicolumn{2}{c}{0.97} \\ \midrule
			\backslashbox{Outcome}{Cluster}                                                					 & Eastern      & Swiss     & Eastern      & Swiss         & Eastern      & Swiss     \\ \hline \\
			\multicolumn{4}{l}{\emph{Simultaneous treatment adoption}} & \\ \\
			\% working in\\ border region                                                                       & 0.073      &  0.995         & 0.091      & 0.622               & 0.184   & \bf{0.016}   \\ \\
			\begin{tabular}[c]{@{}l@{}}\% working\\ in border region,\\  conditional on\\ employment\end{tabular} & 0.083        & 0.823      & 0.133        & 0.930          & 0.189         & \bf{0.019}      \\ \\
			\multicolumn{4}{l}{\emph{Staggered treatment adoption}} & \\ \\
			\% working in\\ border region                                                                       & 0.070      &  0.307         & 0.093      & 0.926                 & 0.195   & \bf{0.014}   \\ \\
			\begin{tabular}[c]{@{}l@{}}\% working in\\ border region,\\  conditional on\\ employment\end{tabular} & 0.090        & 0.966     & 0.129        & 0.767          & 0.179         & \bf{0.017}
			\\ \bottomrule
		\end{tabular}
		\begin{tablenotes}[flushleft]
			\footnotesize
			\item \emph{Notes:} values represent 95\% nonparametric bootstrap $p$-values under the null hypothesis. Values in bold indicate $p$-values under the standard level of statistical significance, $\alpha = 0.05$. 
		\end{tablenotes}
	\end{threeparttable}
\end{table}

\begin{table}[htb]
	\centering
	\begin{threeparttable}	\footnotesize
		\caption{Counterfactual trajectory $p$-values by estimator and cluster.\\ Outcome: \% working in border region, unconditional or conditional on employment.\label{benchmark-pvals}}
		\begin{tabular}{@{}lcccc@{}}
			\toprule
			&  \multicolumn{2}{c}{Eastern}    & \multicolumn{2}{c}{Swiss}    \\
			&  \multicolumn{2}{c}{(2005Q1-2010Q4)}    & \multicolumn{2}{c}{(2005Q1-2008Q4)}  \\ \midrule
			\backslashbox{Estimator}{Outcome} & \multicolumn{1}{c}{Unconditional} & \multicolumn{1}{c}{Conditional} & \multicolumn{1}{c}{Unconditional} & \multicolumn{1}{c}{Conditional} \\  
			\hline \\
			DID           & 0.011 & 0.011	& 0.445 & 0.439  \\
			Matrix completion (ours) & 0.011 & 0.015	& 0.359 & 0.332 \\
			SCM           & 0.009 & 0.005	& 0.841 & 0.982 \\
			\bottomrule
		\end{tabular}
		\begin{tablenotes}[flushleft]
			\footnotesize
			\item \emph{Notes:} values represent 95\% nonparametric bootstrap $p$-values under the null hypothesis, $\hat{\tau}^{LT, pre}=0$.
		\end{tablenotes}
	\end{threeparttable}
\end{table}

\begin{table}[htb]
	\centering
	\begin{threeparttable}	\footnotesize
		\caption{Matrix completion $p$-values by institutional period and cluster.\\ Outcome: \% working in border region, unconditional or conditional on employment. \label{mc-estimates-period-pvals}}
		\begin{tabular}{@{}lcccc@{}}
			\toprule
			&  \multicolumn{2}{c}{Eastern}    & \multicolumn{2}{c}{Swiss}    \\
			&  \multicolumn{2}{c}{(2005Q1-2010Q4)}    & \multicolumn{2}{c}{(2005Q1-2008Q4)}  \\ \midrule
			\backslashbox{Period}{Outcome} & \multicolumn{1}{c}{Unconditional} & \multicolumn{1}{c}{Conditional} & \multicolumn{1}{c}{Unconditional} & \multicolumn{1}{c}{Conditional} \\ 
			\hline \\
			No Treatment            & 0.024  & 0.014 &	0.207 & 0.186 \\
			Single Treatment         & 0.015 & 0.010  & 0.165 & 0.603 \\
			(East.: Schengen; Swiss: FoM)      &  &  &		&   \\
			\bottomrule
		\end{tabular}
		\begin{tablenotes}[flushleft]
			\footnotesize
			\item \emph{Notes:} values represent 95\% nonparametric bootstrap $p$-values under the null hypothesis. See notes to Table \ref{mc-estimates-period}.
		\end{tablenotes}
	\end{threeparttable}
\end{table}

\begin{figure}[htb]
	\centering
	\begin{subfigure}[b]{0.49\textwidth}
	\includegraphics[width=\textwidth]{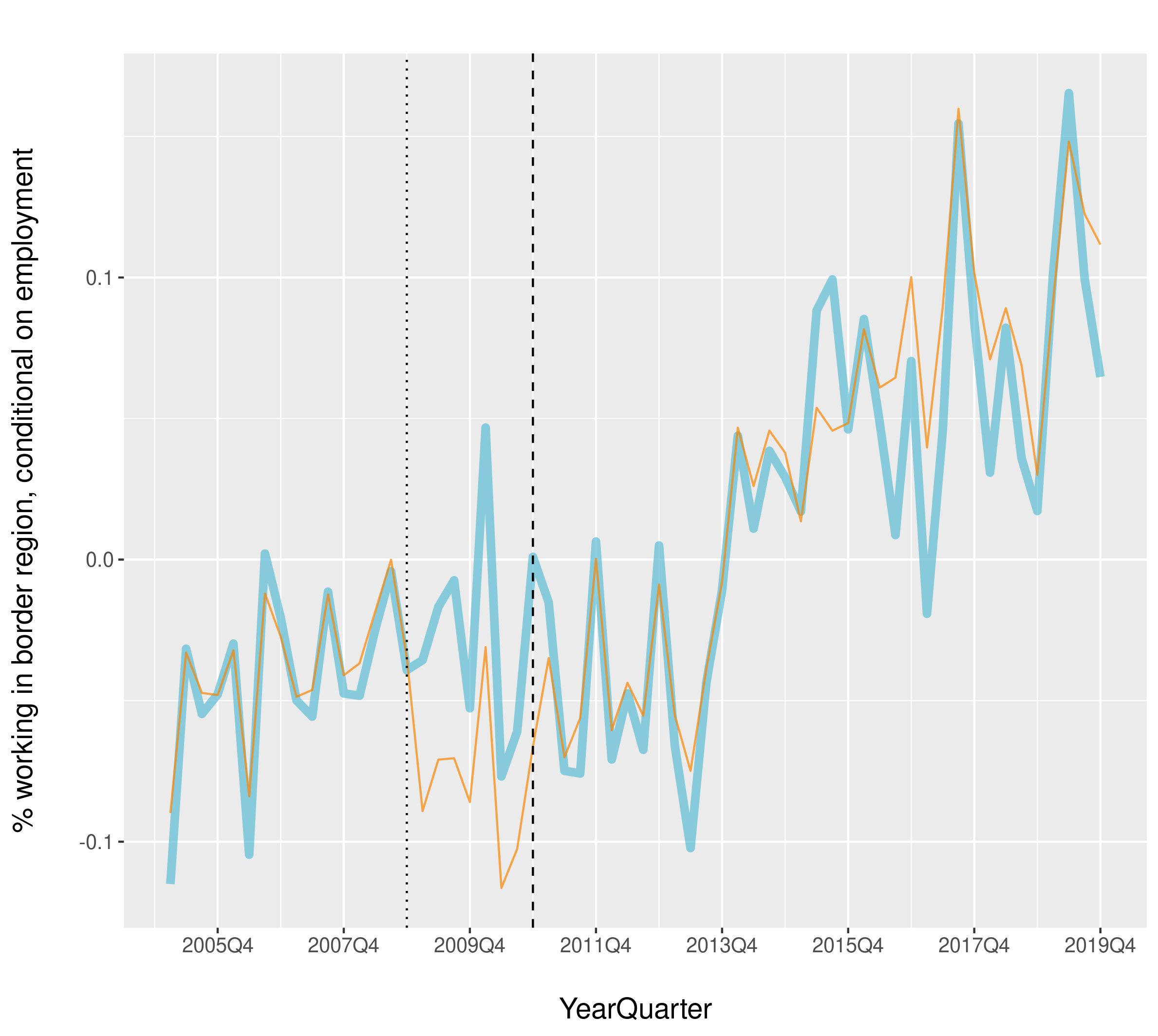}
			\caption{Dominant (first) latent trend}
	\label{fig:mc-trend-first-CBWbordEMPL-covars}
		\end{subfigure}
	\hfill
	\begin{subfigure}[b]{0.49\textwidth}
	\centering
	\includegraphics[width=\textwidth]{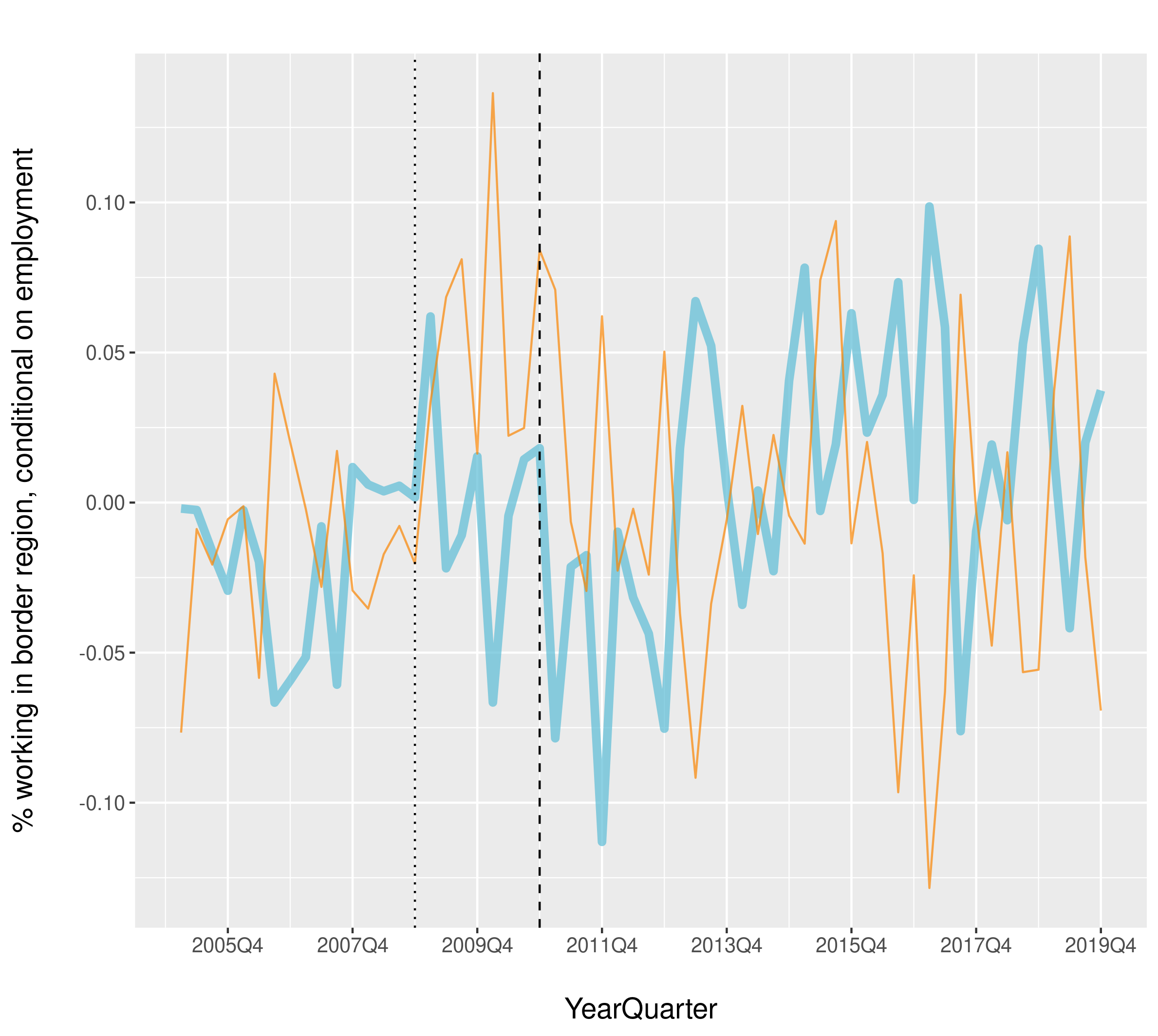}
	\caption{Second latent trend}
	\label{fig:mc-trend-second-CBWbordEMPL-covars}
	\end{subfigure}
	\caption{Matrix completion estimated dominant (first) and second latent trends. Outcome: share of residents working in another country, conditional on employment. \emph{Note:} the dashed vertical lines represent the initial treatment times of the combined treatment, $T^{(\text{Eastern})}_0 = 2011$ and $T^{(\text{Swiss})}_0 = 2009$. \emph{Key:}		{\color{Darjeeling15}{\sampleline{line width=1mm}}}, Eastern regions;
		{\color{Darjeeling14}{\sampleline{line width=0.25mm}}}, Swiss regions. \label{mc-trend}} 
\end{figure}

\clearpage

\bibliographystyle{rss}
\begin{singlespace}
\begin{footnotesize}
\begin{multicols}{2}
\bibliography{references}
\end{multicols}
\end{footnotesize}
\end{singlespace}

\itemize